# FCC-ee Status and Overview

**F. Zimmermann**[a]

[a] *CERN,*
*1211 Geneva 23, Switzerland*
[b] *Name of Institute,*
*Address, Country*
*E-mail:* frank.zimmermann@cern.ch

ABSTRACT: We report on the Future Circular Collider (FCC) Feasibility Study, the mid-term review in autumn 2023, and the longer-term timeline.

[1] Corresponding author.

**Contents**

**1. FCC Feasibility Study**

We are now two years into the FCC Feasibility Study, which was launched by CERN Council in 2021. This Feasibility Study will prepare the ground for an implementation of the first stage of the FCC integrated project, including tunnel construction, technical infrastructure, and the electron-positron collider (FCC-ee), plus the development of financing and organization models. It will also ensure compatibility with a subsequent hadron collider (FCC-hh) and define R&D directions and time schedule for the high-field magnets required for the latter.

In greater detail, the FCC Feasibility Study pursues the following high-level objectives [1]: (1) to demonstrate the geological, technical, environmental and administrative feasibility of the tunnel and surface areas and optimization of placement and layout of the ring and related infrastructure; (2) to pursue, together with the Host States, the preparatory administrative processes required for a potential project approval to identify and remove any showstopper; (3) to optimize the design of the colliders and their injector chains, supported by R&D to develop the needed key technologies; (4) to elaborate a sustainable operational model for the colliders and experiments in terms of human and financial resource needs, as well as environmental aspects and energy efficiency; (5) to develop a consolidated cost estimate, as well as the funding and organizational models needed to enable the project's technical design completion, implementation and operation; (6) to identify substantial resources from outside CERN's budget for the implementation of the first stage of a possible future project (tunnel and FCC-ee); and (7) to consolidate the physics case and detector concepts for both colliders. The results, together with an updated cost estimate, will be summarized in a Feasibility Study Report to be released at the end of 2025, which will serve as an input to the next update of the European Strategy for Particle Physics.

**2. Recent changes and time line**

As one of the first deliverables for the upcoming Feasibility-Study "mid-term review" scheduled for autumn 2023, the placement of the FCC tunnel and surface sites has been optimized for lowest risk [2]. The resulting optimized layout features only 8 (instead of previously 12) surface sites, with a 4-fold super-periodicity (which would allow for up to four collision points), and it has a reduced circumference of 90.6 km, compared with 97.76 km for the earlier FCC conceptual design, published in 2019.

The next update of the European Strategy for Particle Physics is expected to occur around the year 2027. Assuming this update lends support to the first stage of the FCC integrated project, approval could be obtained in 2028. Tunnel groundbreaking and civil engineering may then begin in 2032, accelerator installation in 2040, and FCC-ee accelerator commissioning in 2045.

## 3. FCC articles in this newsletter

The present issue of the ICFA Newsletter includes six articles that elucidate important recent developments and achievements from various areas of the FCC Feasibility Study. First, L. Bromiley and R. Cunningham [3] are setting the scene by reporting on the ongoing civil engineering studies. The next article [4], by F. Carra et al., explores how the accelerators in the arc tunnels will look like. In particular, it describes the design effort towards an arc half-cell mock-up, and the development of supporting structures for both the booster and the collider rings. The following article [5], by L.H. Zhang, H. Damerau, I. Karpov, and A.L. Vanel, describes the limited possible choices for the FCC ring circumference. It is assumed that either a modified LHC or a new superconducting SPS in the existing SPS tunnel is used as the injector for the FCC-hh hadron collider. The requirements for the RF synchronization between injector and collider then yield only a few favourable values of the circumference. A. Chance, B. Dalena and co-authors present a status update for the design of the high energy booster [6], whose layout had to be matched to the new tunnel placement, as for the collider. The booster design studies currently underway focus on longitudinal stability and robustness with respect to errors. An alternate arc optics is also being considered. The last two articles related to FCC-ee, by A. Abramov and R. Bruce [7], and by M. Behtouei, E. Carideo, M. Zobov and M. Migliorati [8], respectively, cover the evolving design of the FCC-ee collimation system and a first assessment of the collimators' impedance.

## Acknowledgments

I would like to thank the international FCC Collaboration for smoothly advancing the FCC Feasibility Study, and, in particular, M. Benedikt, A. Blondel, J. Gutleber, P. Janot, K. Oide, P. Raimondi, D. Shatilov, and T. Raubenheimer.